\documentclass[useAMS,usenatbib]{mn2e}
\usepackage{times}
\usepackage{graphicx}
\usepackage{bm}% bold math
\let\iint\undefined

\usepackage{amsmath,amssymb}
\usepackage{xspace}
\usepackage[english]{babel}
\def\etal{{\rm et al.}\xspace}

\def\ie{{\rm i.e.}\xspace}
\def\eg{{\rm e.g.}\xspace}

\usepackage{url}
\providecommand{\atan}{\ensuremath{\arctan}\xspace}
\newcommand{\sigq}{\ensuremath{\sigma_q}}
\newcommand{\sigu}{\ensuremath{\sigma_u}}
\providecommand{\half}{\ensuremath{\dfrac{1}{2}}}

\providecommand{\sigopt}{\ensuremath{\tfrac{\sigma}{p_0}}}
\newcommand{\bigO}[1]{\ensuremath{\mathcal{O}(#1)}}

\newcommand{\refeq}[1]{Eq.~(\ref{eq:#1})\xspace}
\newcommand{\intinf}{\int_{-\infty}^{+\infty}}
\newcommand{\pmas}{\ensuremath{\hat p_\mathrm{MAS}}\xspace}
\def\deg{\mbox{$^{\circ}$}}
\newcommand{\ricep}{\ensuremath{f_p}\xspace}
\newcommand{\tsqrt}[1]{\ensuremath{\scriptstyle{\sqrt{#1}}}}
\newcommand{\pmin}{\ensuremath{p_\mathrm{min}}}
\newcommand{\pmax}{\ensuremath{p_\mathrm{max}}}
\newcommand{\E}[1]{\ensuremath{E\left[#1\right]}}
\newcommand{\lag}[1]{\ensuremath{L\hspace{-0.1em}_{#1}}}

\usepackage{color}
\definecolor{gris}{gray}{0.50}

\newcommand{\Izero}[1]{I_0\left(#1\right)}
\usepackage{aas_macros}
\newcommand{\citeg}[1]{\citep[\eg][]{#1}}
\def\bsq#1{%both single quotes
\lq{#1}\rq}

\title[Novel estimator for the polarization amplitude]{A novel estimator of the polarization amplitude from normally distributed Stokes parameters}

\author[S. Plaszczynski \etal]{
S. Plaszczynski$^{1}$, L. Montier$^{2,3}$, F. Levrier$^{4}$ and M. Tristram$^{1}$\\
$^1${Laboratoire de l'Acc\'{e}l\'{e}rateur Lin\'{e}aire, Universit\'{e} Paris-Sud 11, CNRS/IN2P3, Orsay, France}\\
$^2${Universit\'{e} de Toulouse, UPS-OMP, IRAP, F-31028 Toulouse cedex 4, France}\\
$^3${CNRS, IRAP, 9 Av. colonel Roche, BP 44346, F-31028 Toulouse cedex 4, France}\\
$^4${LERMA/LRA, ENS Paris et Observatoire de Paris, UMR 8112 du CNRS, 24 rue Lhomond, 75231 Paris cedex 05}\\
}
\date{\today}

\begin{document}

\maketitle

\begin{abstract}
We propose a novel estimator of the polarization amplitude from a    
single measurement of its normally distributed $(Q,U)$ Stokes
components. Based on the properties of
the Rice distribution and dubbed \bsq{MAS} (Modified ASymptotic),
it meets several desirable criteria:
(i) its values lie in the whole positive region; (ii)
its distribution is continuous; (iii) it transforms smoothly with the
signal-to-noise ratio (SNR) from a
Rayleigh-like shape to a Gaussian one; (iv) it is unbiased
and reaches its  components' variance as soon as the SNR exceeds 2; (v) it is analytic 
and can therefore be used on large data-sets.
We also revisit the construction of its associated confidence
intervals and show how the Feldman-Cousins prescription 
efficiently solves the issue of classical intervals lying entirely in the
unphysical negative domain. Such intervals can be used to identify
statistically significant polarized regions
and conversely build masks for
polarization data.
We then consider the case of a general $[Q,U]$ covariance matrix and
perform a generalization of the estimator that preserves its asymptotic
properties. We show that its bias does not depend on the true
polarization angle, and provide an analytic estimate of its variance.
The estimator value, together with its variance, 
provide a powerful point-estimate of the true polarization
amplitude that follows an unbiased Gaussian distribution for a SNR as low as 2.
These results can be applied to the much more general case of
transforming any normally distributed 
random variable from Cartesian to polar coordinates.
\end{abstract}

\begin{keywords}
Physical data and processes: polarization --methods: data analysis -- methods: statistical
\end{keywords}

\maketitle

\section{Introduction}

The advent of high precision experiments dedicated to measuring the
radiation polarization on cosmological scale or
exploring the more local properties of our Galaxy, leads us to revisit the
statistical properties of estimators related to the polarization amplitude.
Polarimeters decompose the incoming monochromatic plane wave radiation
into its $(I,Q,U)$ Stokes components \citep{Chandrasekar1950} in the linear case. According to the
scanning strategy of the instrument, repeated measurements are conducted and
combined, which, owing to the Central Limit Theorem, ensures that 
the Stokes parameters follow a Gaussian distribution. 
However the construction of physical models is most naturally
performed in polar coordinates, \ie using the normalized polarization amplitude (or degree) and
angle. More precisely, astrophysicists are interested in the
\bsq{true} degree of polarization $p_0=\tsqrt{q_0^2+u_0^2}$, and angle
$\psi_0=\tfrac{1}{2}\atan \tfrac{u_0}{q_0}$, where
$q_0=Q_0/I_0$, $u_0=U_0/I_0$, 
and the subscript \bsq{0} emphasizes that we are considering true
quantities.
Working with amplitude and angle data helps assessing the underlying physical processes and  
deserves some statistical attention.

Unlike in the angular case where the naive estimate $\hat \psi=\tfrac{1}{2}\atan\tfrac{u}{q}$ is unbiased \citep{Vinokur1965}, getting a
\bsq{correct} point-estimate for the amplitude from a \textit{single} $(q,u)$
measurement is more involved.
The naive estimate $p=\sqrt{q^2+u^2}$ is indeed strongly biased at low SNR, 
since it does not correct for the power of the experimental noise.
Working instead on $p^2$, one can remove this bias \citeg{Gudbjartsson1995}, but the resultant
distribution, a non-central $\chi^2$ one, is extremely skewed for low SNR 
and the unbiasing induces many negative values.
It is sometimes believed that the Maximum Likelihood (ML) estimator is the optimal
solution since it is known to reach the minimum variance bound. But this is
valid only \textit{asymptotically}, \ie in the limit of a large number
of samples. There is
only one case where the ML estimator is optimal for finite samples:
when the parent distribution is of the exponential form
\citeg{James2007}, which is not the case here at least in the low SNR
regime.  When combining several measurements it however still remains a good solution \citep{Taludkar1991,Sijbers1998}.

An estimator often used in cosmology is based on the most-probable value \citep{Wardle1974}.
Its properties together with a set of other standard estimators was 
reviewed in \citet{Simmons1985}. All these estimators are
however \textit{discontinuous}: their distribution is a mixture of a
discrete peak at zero and a positive tail.
While statistically valid, this in practice is very undesirable. Their
bias and risk are small because they are computed in a \textit{ensemble
  average} sense. But an ergodicity argument cannot be invoked since
the user generally works on a single realization of the sky.
In practice when applying these estimators, for instance to a pixelized
map, the user ends up with a large number of zeros and does not know 
how to treat them.
Bayesian estimators that are continuous were proposed by
\citet{Quinn2012}. However, as we will see in Sect. \ref{sec:mas}, their
distribution is very skewed and has a cutoff value.

The aim of this work is to cure these issues and provide a
polarization amplitude estimator from a bi-variate
normally distributed $(q,u)$ measurement that is continuous and lies in
the whole positive region.
We will particularly take care of the overall shape of the estimator distribution, not
only its first two moments as characterized by the bias and risk.

Previous works focused on a $[q,u]$ covariance matrix proportional to
identity, $C=\sigma\mathbf{1}$, what we will call the \textit{canonical} case.
Given the extreme sensitivity of the current and planned 
experiments, we will also consider the case of a general covariance
matrix, \ie including some ellipticity ($\sigq \ne \sigu$) and
correlation ($\rho$):
\begin{equation}
C=
\begin{pmatrix}
  \sigq^2 & \rho \sigq \sigu\\ 
  \rho \sigq \sigu & \sigu^2
\end{pmatrix}.
\end{equation}

In Sect \ref{sec:asymptotic}, we will first review the asymptotic
properties of the naive estimator, in the canonical case of a $[q,u]$
covariance matrix proportional to identity, \ie $\sigq=\sigu=\sigma,\rho=0$. This will allow us to
retrieve the asymptotic estimator and cure its discontinuity while still
keeping rapid convergence to the asymptotic limit. We will
characterize our estimator in Sect. \ref{sec:perf} not only with its
first order moments but with its full distribution for which we will
provide an analytic approximation. When building confidence intervals in Sect.
\ref{sec:cl}, we will cure the classical problem of regions lying into
the unphysical region by applying the Feldman-Cousins prescription. It
will allow us to obtain physical intervals without ever being
\bsq{conservative} (as defined in Sect.\ref{sec:cl}). An analytic description of the interval will be given
for our estimator.
Then in Sect. \ref{sec:general} we will consider the case of a general
$[q,u]$ covariance matrix before concluding that our estimator can be used efficiently
to provide reliable (Gaussian) estimates in regions of SNR above 2, and
conversely construct polarization masks for regions with a low
statistical significance.

\section{Asymptotic properties of the amplitude distribution}
\label{sec:asymptotic}

\subsection{Approximations to the Rice distribution}
\label{sec:canon}

We begin by revisiting the asymptotic properties of the amplitude
distribution in the case where the $(q,u)$ Stokes parameters are
drawn from a Gaussian centred around the true values ($q_0,u_0)$ and
with a simple covariance matrix proportional to the identity ($\sigq=\sigu=\sigma$).

The change of $(q,u)$ variables into polar coordinates
\footnote{Throughout the text we will work with the angular polar
  coordinates $\phi$,
  keeping in mind that the polarization angle, which is a spin-2
  quantity, is defined by $\psi=\phi/2$. The \atan function is
  classically generalized to span the whole $[-\pi,\pi]$ range.}
\begin{equation}
  \label{eq:polar}
  \begin{split}
    p&=\sqrt{q^2+u^2},\\
    \phi&=\atan\dfrac{u}{q},
  \end{split}
\end{equation}
leads to the bi-variate polar distribution:
\begin{equation}
  \label{eq:rice2d}
  f_{p,\phi}(p,\phi)= \dfrac{p}{2\pi\sigma^2} e^{-\dfrac{p^2+p_0^2}{2\sigma^2}}e^{\dfrac{p p_0\cos(\phi-\phi_0)}{\sigma^2}},
\end{equation}
where we have introduced the true polar values:
\begin{equation}
  \begin{split}
    p_0&=\sqrt{q_0^2+u_0^2},\\
    \phi_0&=\atan\dfrac{u_0}{q_0}.
  \end{split}
\end{equation}

Our aim is then to estimate the true amplitude $p_0$ and angle $\phi_0$. 
Marginalization over the angle leads to the Rice distribution
\citep{Rice1945} that does not depend anymore on the true $\phi_0$ value:
\begin{equation}
  \label{eq:rice}
  \ricep(p)=\dfrac{p}{\sigma^2}
  e^{-\dfrac{p^2+p_0^2}{2\sigma^2}} \Izero{\dfrac{p p_0}{\sigma^2}},
\end{equation}
where $I_0$ denotes the modified Bessel function of order 0.
Its moments can be computed exactly using
\citet{GradshteynRyzhik2007} Eq.~(6.631), $I_0(z)=J_0(i z)$ and the connection
between Kummer's confluent hypergeometric function (noted $_1F_1$ or
$M$) and the Laguerre polynomials \lag{k}
\citep[][Eq.~(18.11.2)]{NIST2010}, which gives:
\begin{align}
  \label{eq:ricemom}
    \E{p}&=\sqrt{\dfrac{\pi}{2}} \sigma \lag{\frac{1}{2}}\left(-\dfrac{p_0^2}{2\sigma^2}\right),\\
    \E{p^2}&=2\sigma^2 + p_0^2,
\end{align}
where the half-order Laguerre polynomial $\lag{\frac{1}{2}}$ can be conveniently computed from:
\begin{equation}
  \label{eq:laghalf}
\lag{\frac{1}{2}}(z)=e^{z/2}\left( (1-z)I_0(-z/2)-z I_1(-z/2) \right).
\end{equation}

The moments allow us to build the variance $\E{p^2}-\E{p}^2$ and the risk
$=\E{(p-p_0)^2}$, which depends on the true  $p_0$ value.
For a large SNR, \ie when $\epsilon\equiv\dfrac{\sigma}{p_0} \to 0$,
the leading order expansion of the mean is:
\begin{align}
  \label{eq:mean}
  \E{p}&=p_0(1+\epsilon^2/2) +\bigO{\epsilon^4},\nonumber \\
  &=p_0+\dfrac{\sigma^2}{2p_0}+\bigO{\epsilon^4},
\end{align}
while, to same order, the variance is:  
\begin{equation}
  \label{eq:variance}
  V(p)=\sigma^2+\bigO{\epsilon^4}.
\end{equation}
The mean and variance both involve the Gaussian variance $\sigma^2$. To avoid
confusion in the following, we will denote its first meaning as a
(non-linear) \bsq{noise-bias} and call it $b^2$.

It is often claimed \citeg{Gudbjartsson1995,SijbersThesis1998,Cardenas2008} that the Rice
distribution converges asymptotically to a Gaussian: 
\begin{equation}
  \label{eq:gaussapprox}
f_p\to \mathcal{N}(\sqrt{p_0^2+\sigma^2},\sigma^2),
  \end{equation}
where $\mathcal{N}(\mu,\sigma^2)$ denotes a Gaussian distribution of
mean $\mu$ and variance $\sigma^2$.

\begin{figure}
  \centering
\includegraphics[width=.45\textwidth]{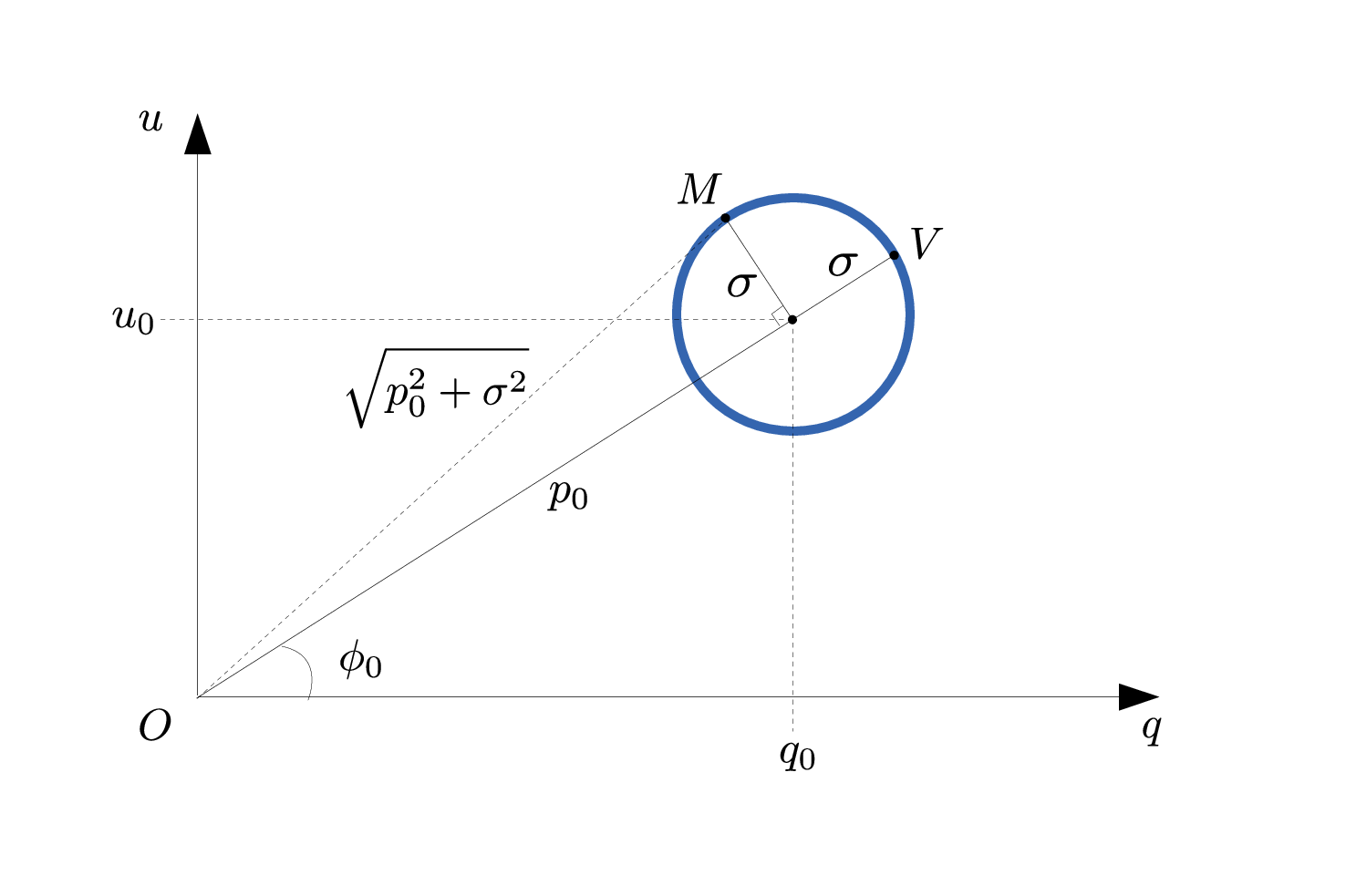}
\caption{\label{fig:geo_simple} Illustration of the mean and variance
  of the amplitude distribution in the canonical case from a sampling
  point of view. $(q,u)$ samples are drawn according to a Gaussian of
  mean $(q_0,u_0)$ and variance $\sigma$. The circle represents the
  1-$\sigma$ iso-probability contour. One considers the distance to
  the origin of samples located uniformly on that circle. In the asymptotic
  case, \ie when the circle is far from the origin, the distance
  distribution  is (almost) symmetric around the value corresponding
  to that of the $M$ point, 
  which is \textit{orthogonal} to the direction
  towards the circle centre. The mean value there is
  ${\tsqrt{p_0^2+\sigma^2}}$. The distribution lies in the $p_0\pm\sigma$
range and has a variance of $\sigma$ estimated \textit{along} the direction to
the centre. By considering the angular distribution of the samples,
one also finds that it is centred on $\phi_0$ (\ie unbiased) and has
a deviation of $\tfrac{\sigma}{p_0}$, as confirmed by a direct calculation \citep{Vinokur1965}. This construction is only
approximate, but captures the essentials of the mean and variance computations.
}
\end{figure}

The origin of these values for the  mean and variance can be
understood from the simple geometric construction of Fig.~\ref{fig:geo_simple}.
That the distribution converges to a Gaussian one is, as far as we
know, not justified in the literature so we re-examine that statement in
some detail. 

\begin{figure}
  \centering
  \includegraphics[width=.5\textwidth]{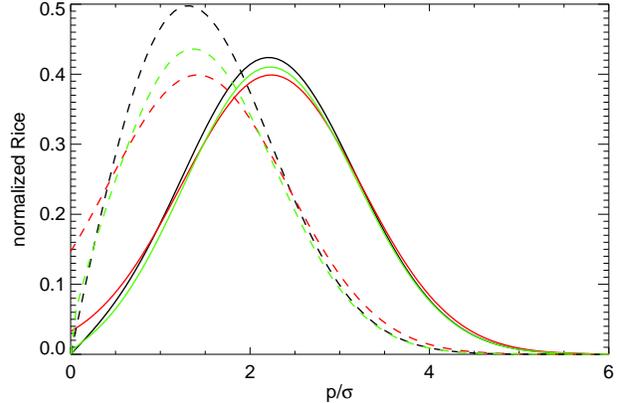}
\caption{Approximations to the Rice distribution for $p_0/\sigma=2$
  (solid lines) and $p_0/\sigma=1$ (dashed lines). The black curves
  correspond to the exact Rice scaled distribution, the red ones to the
  traditional Gaussian approximation (\refeq{gaussapprox}),  and
  the green ones to our \refeq{riceapprox} approximation.}
\label{fig:riceapprox}
\end{figure}

For a large argument, the modified Bessel function converges to \citep[][Eq.~(10.40.1)]{NIST2010}: 
\begin{equation}
  \label{eq:I0expansion}
  \Izero{z} \to \dfrac{e^{z}}{\sqrt{2\pi z}},
\end{equation}
and then the Rice distribution to:
\begin{equation}
 \label{eq:riceapprox}
  f_p \rightarrow \sqrt{\dfrac{p}{p_0}}{\cal N}(p_0,\sigma^2).
\end{equation}
This approximation is valid for a SNR above about 1 (see Fig.~\ref{fig:riceapprox}).   

This distribution then converges to a Gaussian, for a SNR larger than
about 2, as shown on Fig.~\ref{fig:riceapprox}.
The reason can be understood by making the change of variable
$p^\prime=\tfrac{p-p_0}{\sigma}$ and expanding the square-root to
first order in \sigopt, the distribution of the scaled variable tends to: 
\begin{equation}
 f_{p^\prime} \rightarrow \mathcal{N}(0,1)+\dfrac{\sigma}{2 p_0}p^\prime\mathcal{N}(0,1),
\end{equation}
which exhibits a corrective term to a pure Gaussian that is getting smaller with
$\epsilon=\dfrac{\sigma}{p_0}$. It can then be verified that
this approximation leads indeed to the two moments of \refeq{mean} and \refeq{variance}. 

The first order effect of the corrective term can thus be captured into a
bias of the Gaussian mean which converges to
\refeq{mean}. Up to \textit{first order} this is indeed the Taylor
expansion of $\sqrt{p_0^2+\sigma^2}$. However the next order term in
this expansion is negative ($-\tfrac{1}{8}\sigma^4/p_0^3$), while the one from
the exact mean expression is positive ($+\tfrac{1}{8}\sigma^4/p_0^3$).
It is therefore more correct to use simply $p_0+\tfrac{\sigma^2}{2p_0}$ for the Gaussian mean.

What we learned so far, is that the ${\cal  N}(\sqrt{p_0^2+\sigma^2},\sigma^2)$ Rice
approximation is a first-order asymptotic expansion valid
for $p_0/\sigma\gtrsim 2$. A slightly better approximation is obtained
from the first-order expansion of the mean,
$\mathcal{N}(p_0+\tfrac{\sigma^2}{2p_0},\sigma^2)$,  
and yet a better one by $\sqrt{\tfrac{p}{p_0}}{\cal N}(p_0,\sigma^2)$, which 
is valid above $p_0/\sigma\gtrsim 1$.

\subsection{Modified ASymptotic estimator (MAS)}
\label{sec:mas}
We now address the question of building an estimator of the true $p_0$
value with \bsq{good} properties, which is a somewhat subjective
notion. We feel however that an essential property is convergence as
fast as possible with the SNR to the true
value but also that the estimator distribution has a \bsq{reasonable}
shape (this will be clarified later). 
Keeping in mind that building a perfectly unbiased estimator for a very low $p_0$
is mathematically impossible (see Appendix \ref{app:A}), we will
focus on the asymptotic approximations to the Rice distribution. 
To avoid confusion in the following, we will add an index \bsq{i} to the measurement,
even-though we are considering a single sample.

We are looking for a \bsq{satisfactory} estimator given a single sample $p_i=\sqrt{q_i^2+u_i^2}$.
From the standard Rice approximation $\mathcal{N}(\sqrt{p_0^2+\sigma^2},\sigma^2)$, the
maximum likelihood estimator in this case is straightforwardly: 
\begin{equation}
  \label{eq:as}
  \hat p_{AS}=\sqrt{p_i^2-\sigma^2}.
\end{equation}
Using our slightly more precise approximation
$\mathcal{N}(p_0+\tfrac{\sigma^2}{2p_0},\sigma^2)$ one obtains:
\begin{equation}
  \label{eq:as2}
  \hat p_{AS^\prime}=\half (p_i+\sqrt{p_i^2-2\sigma^2}),
\end{equation}
which is also the ML estimator using our most precise approximation
$\sqrt{\tfrac{p}{p_0}}\mathcal{N}(p_0,\sigma^2)$.

In this form we encounter the problem of dealing with negative values 
under the square-root as discussed in the introduction.
We show how to build a simple continuous analytic
estimator that expands in the whole positive region, and converges
rapidly to the asymptotic limit.
The first order expansion of both \refeq{as} and \refeq{as2} is
\begin{equation}
  \label{eq:mas1}
  \hat p=p_i-\dfrac{\sigma^2}{2p_i},
\end{equation}
which is also the most probable estimator of our
$\sqrt{\tfrac{p}{p_0}}\mathcal{N}(p_0,\sigma^2)$ approximation.
This estimator diverges for low values. We want to modify it
based on the following requirements:
\begin{enumerate}
\item the transformation must be smooth, in order to avoid Jacobian peak effects,
\item it must converge to the asymptotic result (\refeq{mas1}) for a SNR around 2,
\item the samples must always remain positive,
\item the estimator distribution transforms smoothly to an unbiased
  Gaussian as the SNR increases.
\end{enumerate}

We then consider transformations of the form:
\begin{equation}
  \hat p =p_i- \sigma^2\frac{1-e^{-\lambda p_i^2 / \sigma^2}}{2p_i},
\end{equation}
where $\lambda>0$ is to be discussed, which preserves the correct asymptotic
limit while converging linearly to 0 for low values:
\begin{equation}
  \label{eq:lim}
\hat p =\left(1-\dfrac{\lambda}{2}\right) p_i +\bigO{p_i^2}.
\end{equation}

In order to fulfill (ii) we wish $\lambda \ge 1$.
On the other hand, $\lambda$ should not exceed $2$ since
otherwise the derivative around 0 would
become negative (see \refeq{lim}) and we would fail (iii).
For $\lambda$ around 2, the estimator distribution is peaked
at 0 and similar to an exponential. When transforming to a Gaussian
with the SNR, it develops an intermediate minimum that complicates its
overall shape.
In contrast, for $\lambda$ around 1, the distribution transforms
from a Rayleigh-like one to a Gaussian one without introducing
a secondary extremum, which is
similar to the Rice case and will be further discussed in Sect. \ref{sec:distrib}. Given the marginal gain of using
$\lambda=2$ and its induced complexity on the distribution, we consider $\lambda=1$ as our optimal solution.

We then propose the following Modified ASymptotic (MAS) estimator:
\begin{equation}
  \label{eq:mas}
  \pmas=p_i- \sigma^2\frac{1-e^{-p_i^2 / \sigma^2}}{2p_i}.
\end{equation}

\begin{figure}
  \centering
  \includegraphics[width=.5\textwidth]{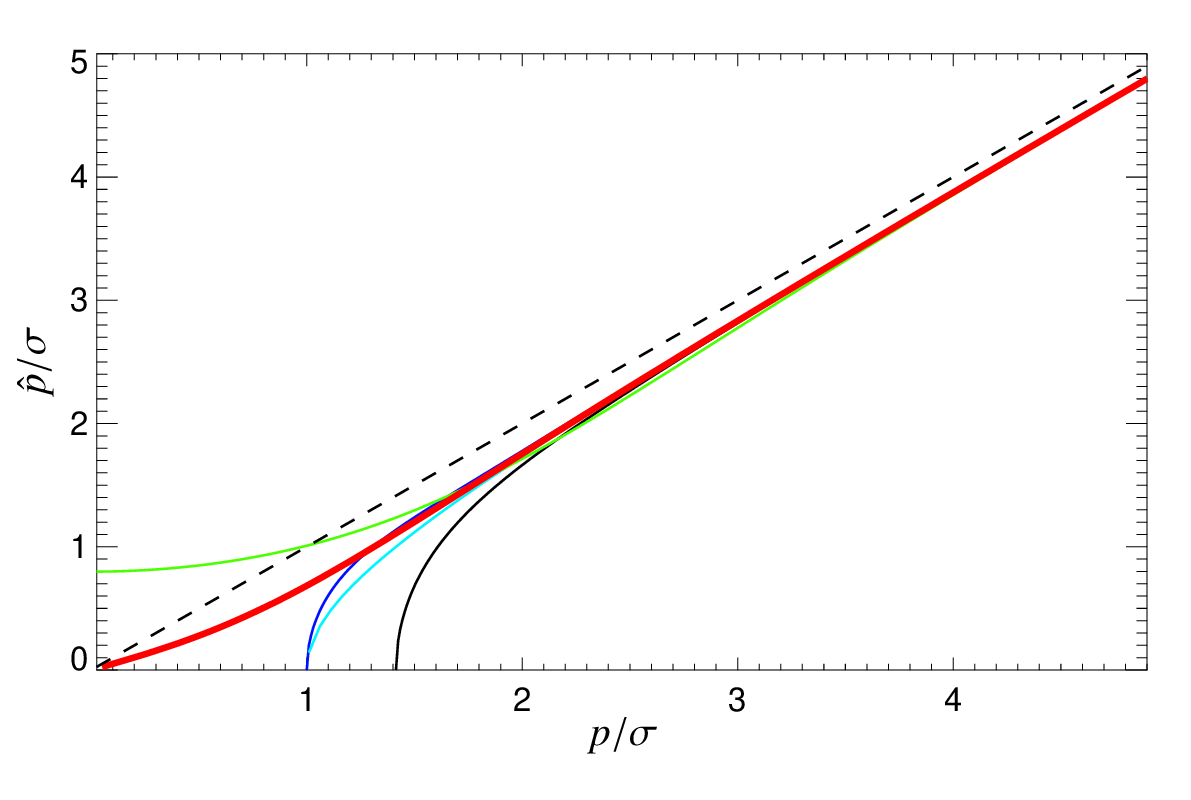}
\caption{\label{fig:curve} Transformation curve of the MAS
  estimator (in red). We also show some other classical
  estimator curves: in light-blue, the Asymptotic (\refeq{as}), in blue
  the Most Probable \citep{Wardle1974} and in black
  the Maximum Likelihood \citep{Simmons1985}. They are discontinuous and the latter
  two non-analytic. Also shown in green is the curve of the
  posterior-mean Bayesian estimator \citep{Quinn2012} with
  a uniform prior on $p_0/\sigma$.
 The dashed line represents the naive estimator.} 
\end{figure}

We show on Fig.~\ref{fig:curve} its transformation curve, together with some
other classical estimators, demonstrating how it extrapolates smoothly from the
asymptotic regime down to 0. This figure reveals that:
\begin{itemize}
\item the Most Probable estimator \citep{Wardle1974} has essentially the same
  properties as the simple asymptotic one of \refeq{as};
\item these two, together with the ML one \citep{Simmons1985}, are discontinuous, \ie have
  a non-differentiable transform at one point which leads to a set of
  discrete samples at 0;
\item the one-dimensional posterior-mean Bayesian estimator with a
  uniform prior in $p_0$
  is lower-bounded at $\tsqrt{\tfrac{2}{\pi}}$$\simeq 0.8 \times\sigma$ which can be
  verified from its expression that is analytic in the moderate SNR regime:
\footnote{The analytic computation is performed after a change of
  variable into the scaled (SNR) variable and  letting $1/\sigma \to
  \infty$. The results holds up to very high polarization values.}
  \begin{align}
    \hat p_\mathrm{mean}&=\dfrac{\int_0^1 p_0 f_p(p|p_0) dp_0}{\int_0^1
      f_p(p|p_0) dp_0} \nonumber \\
    & \simeq \left [ \tfrac{1}{\sigma} \sqrt{\tfrac{\pi}{2}}
      e^{-\tfrac{p^2}{4\sigma^2}} \Izero{\tfrac{p^2}{4\sigma^2}} \right]^{-1}.
  \end{align}
Furthermore, such curves that have a null derivative at low SNR, which is the case
of all Bayesian estimators presented in \citet{Quinn2012}, lead to
extremely skewed distribution at low SNR as can be inferred from
transforming samples drawn from a Rayleigh-type distribution along the $p/\sigma$ axis.
\item all these estimators but the naive one have the correct asymptotic
  limit (which is \refeq{mean}) and differ by the way they behave at low values.
\end{itemize}

%%%%%%%%%%%%%%%%%%%%%%%%%%%%%%%%%%%%%%%%%%%%%%%%%%%%%%%%%%%%%%%%%
\section{Performance of the MAS estimator}
\label{sec:perf}

\subsection{Distribution}
\label{sec:distrib}

\begin{figure}
  \centering
  \includegraphics[width=.5\textwidth]{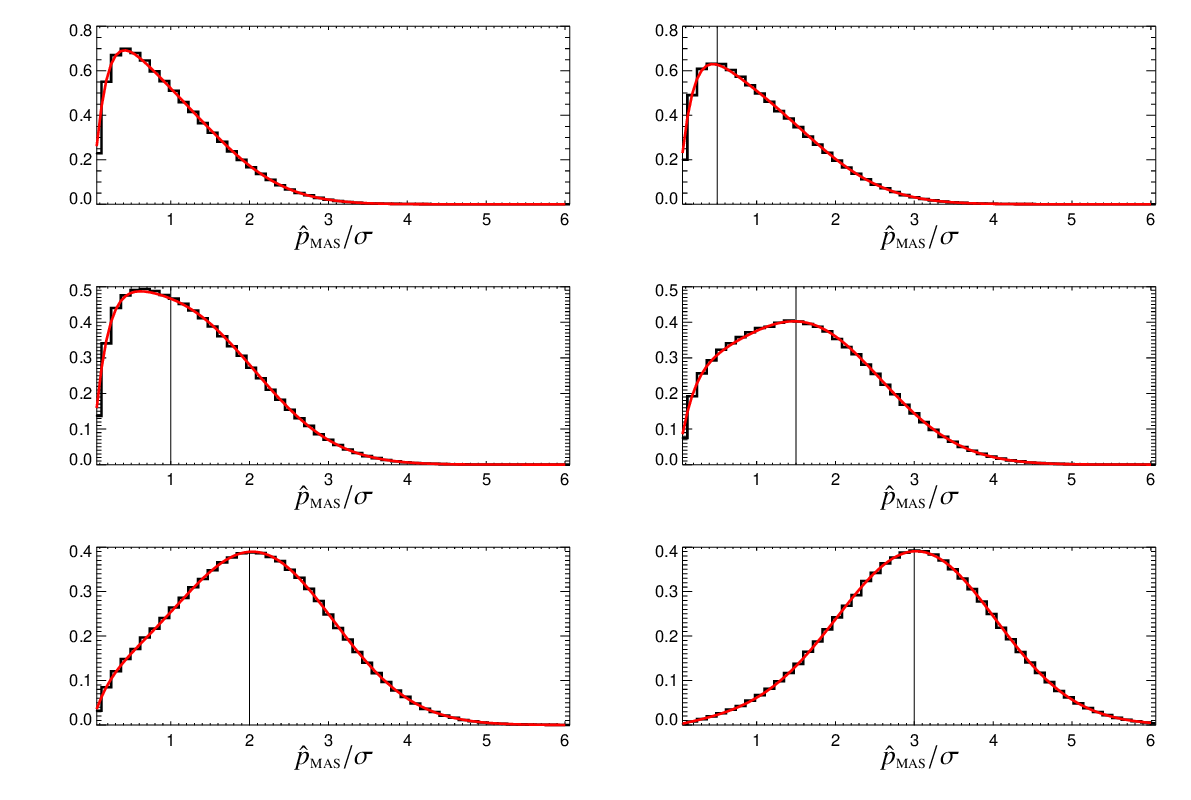}
\caption{\label{fig:pmas_distrib}
MAS estimator distribution in the canonical case, as
  obtained from the Monte-Carlo simulations, for several $p_0/\sigma$ values (shown
  as the vertical line). From left to right and top to bottom $p_0=0,0.5,1,1.5,2,3$.
  The analytic approximation discussed in the text
  (\refeq{pmas_distrib}) is superimposed in red.} 
\end{figure}

We study the distribution of the MAS estimator \refeq{mas}, in the
canonical case, using Monte-Carlo simulations. For a given $p_0$
value, we shoot $10^{6}~ (q_i,u_i)$ normally distributed samples
centred on $q_0=p_0 \cos \phi_0, u_0=p_0 \sin\phi_0$, where $\phi_0$ is drawn from a uniform
distribution on $[-\pi,\pi]$. We then compute $p_i=\sqrt{q_i^2+u_i^2}$,
transform the samples according to \refeq{mas} and project them into a
histogram in order to obtain the probability density function.
Fig.~\ref{fig:pmas_distrib} shows some distributions for increasing
$p_0$ values which exhibit how they change smoothly from
Rayleigh-like at low SNR to Gaussian as soon as $p_0/\sigma \gtrsim 2$.

We work out in the following an analytic description of its
distribution, which is useful for implementing a likelihood function.
Using the scaled variable $p\leftarrow \tfrac{p}{\sigma},
p_0\leftarrow \tfrac{p_0}{\sigma}$, the MAS transformation reads,
dropping out the \bsq{i} subscript:
\begin{equation}
  \hat p=p-\dfrac{1-e^{-p^2}}{2p}.
\end{equation}
The standard rules of random variable transformation requires inverting
this equation which does not have an exact analytic expression. We note however
that in the asymptotic limit, the exponential can be neglected and the
inverse is $p =\tfrac{1}{2} (\hat p+\sqrt{\hat p^2+2})$. From numerical
comparison to the inverse, we find it sufficient to complement it with an
exponential. We obtain the following approximate inverse relation:
\begin{equation}
  \label{eq:pmas_inv}
   p\simeq g(\hat p)=\half (\hat p+\sqrt{\hat p^2+2})(1-e^{-a \hat p}),
\end{equation}
with $a=3.17$. This approximation is valid in the whole positive range below the percent level. 

The distribution of the $\hat p$ estimator is then obtained from the
transformation of the Rice distribution \ricep (\refeq{rice}) as:
%%%changer R partout
\begin{align}
  \label{eq:pmas_distrib}
  f_{\hat p}(p)&=g^\prime(p) \ricep(g(p)) \nonumber \\
  & \simeq \dfrac{(p + \sqrt{p^2 + 2})(a\sqrt{p^2 + 2} + e^{a p}
    -1)}{2e^{ap}\sqrt{p^2 + 2}} \nonumber \\
  & \quad \times \ricep\left(\half (p+\sqrt{p^2+2})(1-e^{-a p})\right),
\end{align}
and the complete distribution is given by $f_{\hat p}\left(\tfrac{p}{\sigma}\right)/\sigma$.
This analytic approximation is excellent as shown for some examples on Fig.~\ref{fig:pmas_distrib}.

\subsection{Bias and risk}

\begin{figure}
  \centering
  \includegraphics[width=.5\textwidth]{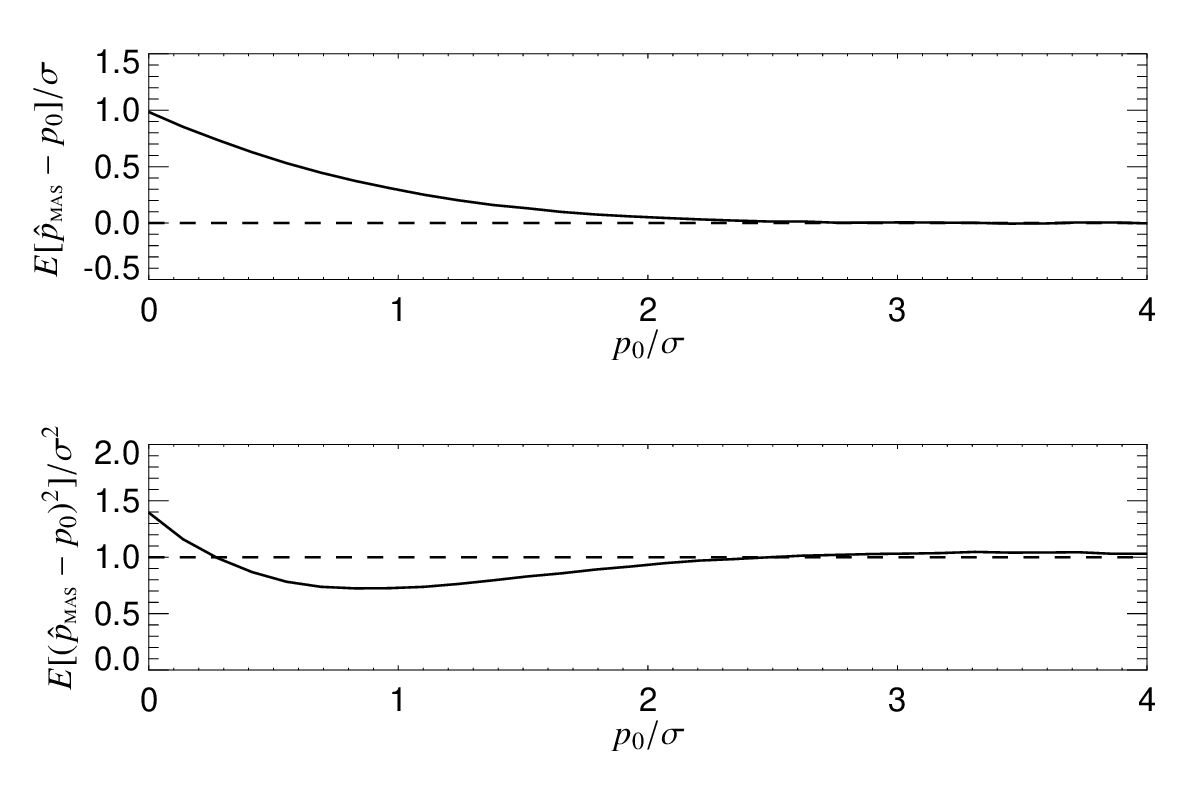}
\caption{\label{fig:bias_risk} Estimate of the normalized bias (top)
  and risk (bottom) of
  the modified asymptotic estimator (MAS) in the canonical case, as obtained
  from Monte-Carlo simulations.}
\end{figure}

The first two orders of the estimator statistics are characterized by the
normalized bias $E[\pmas-p_0]/\sigma$ and risk $E[(\pmas-p_0)^2]/\sigma^2$, using
Monte-Carlo simulations. They are shown on Fig.~\ref{fig:bias_risk}.
For a SNR as low as 2, the estimator is essentially unbiased and has
a $\sigma^2$ risk.

\subsection{Confidence intervals}
\label{sec:cl}

\begin{figure}
  \centering
  \includegraphics[width=.5\textwidth]{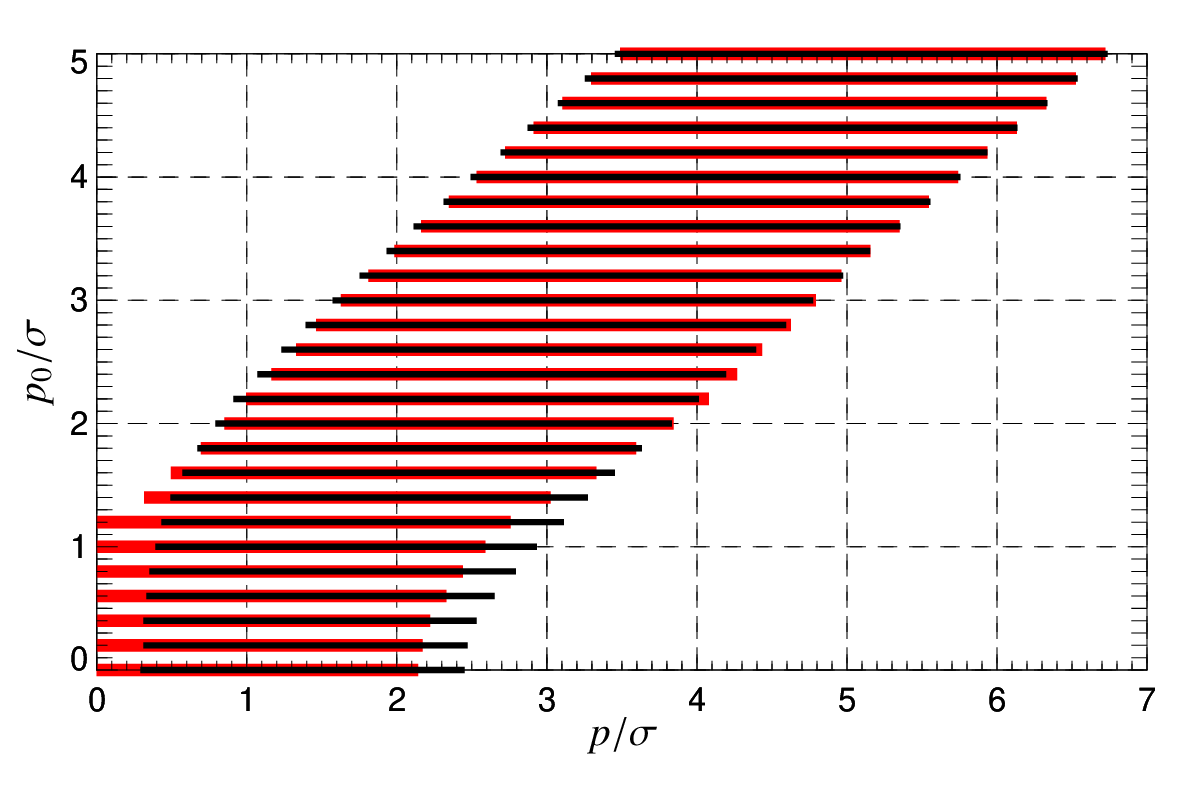}
\caption{\label{fig:rice_cl} Construction of a 90\% CL interval
  for the naive estimator in the canonical case, using the central
  confidence region (black lines) or the Feldman-Cousins
  prescription (red lines). For a measured
  sample value $p/\sigma$ one reads off the associated confidence interval
  on the vertical axis. For low values ($p/\sigma < 0.29$) 
  the central interval lies entirely inside the unphysical negative
  region. This is cured by applying the
  Feldman-Cousins prescription.}
\end{figure}

We emphasize that the characterization of estimators in
terms of their mean and risk may lead to over-simplification
and misunderstandings in a community accustomed to considering a number
with an \bsq{error} as originating from a Gaussian distribution.
Instead, providing a confidence interval at some given significance level $\alpha$ is
more complete since it is independent of the shape of the estimator distribution.
The construction of a classical confidence interval
is an old and well defined statistical procedure
\citep{Neyman1937}. It however does not specify 
uniquely the acceptance region, since
given some fixed $p_0$ value $Pr(\hat p\in [\pmin,\pmax]|p_0)=\alpha$
is insufficient to fix the interval.
One must choose an additional free criterion.
A common choice is to use the central confidence interval:
\begin{equation}
  \label{eq:central}
  Pr(\hat p < \pmin)=Pr(\hat p > \pmax) =\dfrac{1-\alpha}{2}.
\end{equation}
It may however lead to the situation of providing
an \bsq{empty-set} $\{0\}$ or, equivalently, an interval lying entirely in the
unphysical region (see Fig.~\ref{fig:rice_cl}), 
which is statistically valid, but uncomfortable to an analyst.
One solution is to enlarge the interval given an arbitrary construction
to provide a \bsq{conservative} one, a procedure already used for the naive estimator
\citep{Simmons1985}.

Here we rather advocate using the Feldman-Cousins
prescription \citep{FC}, which, for the free criterion, 
uses an ordering of the likelihood ratios. The authors showed that the problem
of empty-sets relates to intervals failing a goodness-of-fit
test. Their procedure naturally decouples this test from the
construction of the interval, effectively removing the empty-set issue
without ever being conservative.
We show in the following how to perform it in our case.

We consider some estimator $\hat p$ for which we can compute the
distribution $\hat f(p | p_0)$, possibly via Monte-Carlo simulation.
We pre-compute first its maximum likelihood curve, \ie the $p_0$ value
for which $\hat f(p | p_0)$ is maximum.
We then \textit{scan} $p_0$ values, and at each step:
\begin{enumerate}
\item compute the likelihood ratio curve as
  $R(p)=\dfrac{\hat f(p|p_0)}{\hat f(p|p_{ML})}$, where $p_{ML}(p)$ is obtained
  from our pre-computation,
\item solve numerically the system
$\begin{cases}
R(\pmin)=R(\pmax) \\ 
\int_{\pmin}^{\pmax} dp \hat f(p|p_0)=\alpha\\
\end{cases}$,
\item report the $[\pmin,\pmax]$ interval  for this $p_0$ value horizontally
  on a graph  known as the \bsq{confidence belt} \citep[e.g.][see
  also Fig.~\ref{fig:rice_cl}]{PDG2012}.
\end{enumerate}
The standard Neyman's inversion statement then allows, for a given
 $p/\sigma$ sample, to measure its $\alpha$-level confidence interval
on the vertical axis.

We show on Fig.~\ref{fig:rice_cl} the result of this computation
at the $\alpha=0.90$ level for the Rice distribution (\ie the naive estimator) and
compare the limits obtained to the classical central intervals.
The empty-set at low $p/\sigma$ values indeed disappears and the user can
now report a confidence intervals for any measured value, without ever being
conservative.
Asymptotically ($p/\sigma \gtrsim 3.5$) both constructions agree, but we
obtain tighter constraints in the intermediate region $p/\sigma\in[2.2,3.5]$. 

\begin{figure}
  \centering
  \includegraphics[width=.5\textwidth]{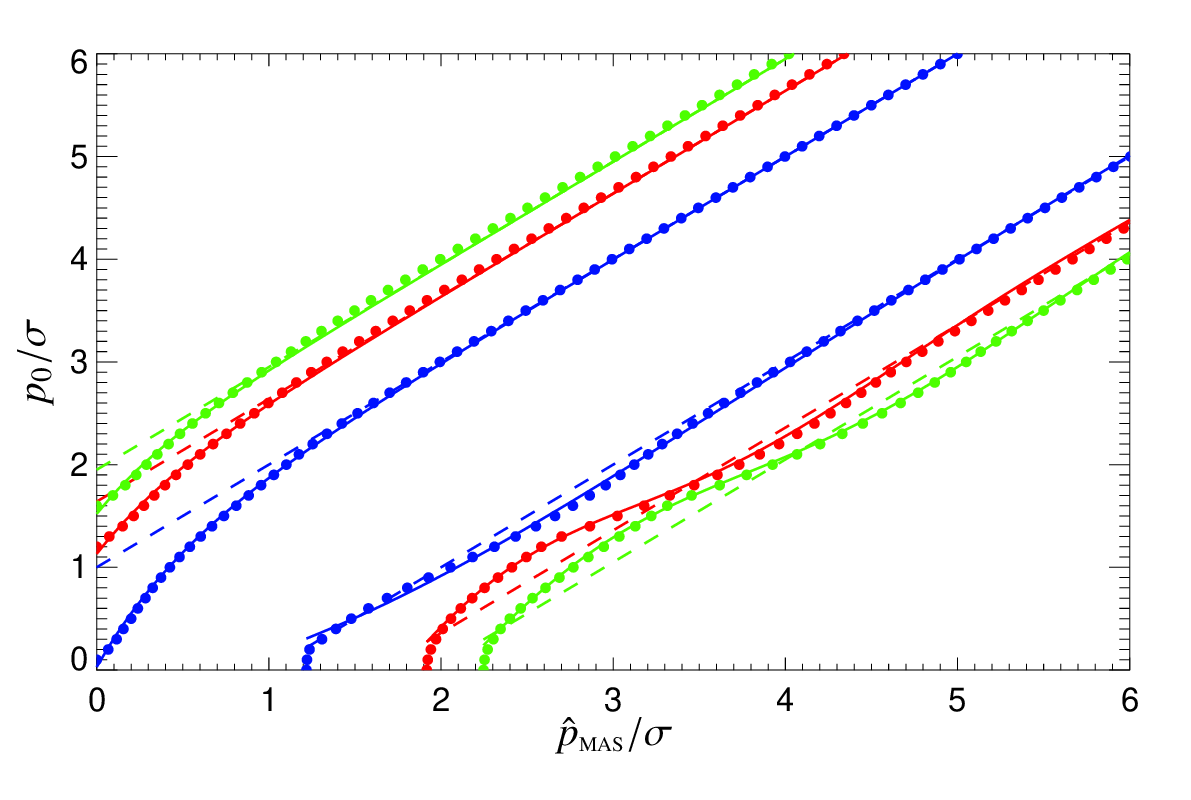}
\caption{\label{fig:pmas_cl} Confidence belts of the normalized
  MAS estimator, using the Feldman-Cousins prescription, for 0.68
  (blue dots) ,
  0.90 (red dots) and 0.95 (green dots) confidence levels. The dashed lines
  correspond to the Gaussian intervals that are reached
  asymptotically. The solid lines correspond to the analytic description
  provided in the text. For a given $\pmas/\sigma$ value, the
  corresponding confidence interval is read vertically.}
\end{figure}

Using the Feldman-Cousins prescription, we then build the confidence belts of the MAS estimator at
the 0.68, 0.90 and 0.95 confidence levels. They are shown on
Fig.~\ref{fig:pmas_cl}. For convenience we provide the following
analytic approximations to the scaled upper and lower limits at the  $\alpha$
significance level:
\begin{equation}
\label{eq:approxcls}
\begin{split}
  \pmin^\alpha&=\pmas -p_\alpha(1+\beta e^{-\gamma \pmas}\sin(\omega \pmas+\phi)), \\ 
  \pmax^\alpha&=\pmas+p_\alpha(1-\beta e^{-\gamma \pmas}),
\end{split}
\end{equation}
where $p_\alpha=\sqrt{2} \mathrm{Erf}^{-1}(\alpha)$ is the $\alpha$-point of the
Gaussian distribution that is reached
asymptotically, and the parameters are given in Table \ref{tab:clparams} for the three significance levels.

\begin{table}
  \centering
  \begin{tabular}{ccccccc}
\hline\hline
    Bound & $\alpha$ & $p_\alpha$ & $\beta$ & $\gamma$ & $\omega$ &
    $\phi$ \\
    \hline
    $\pmin$ & 0.68 & 1 & 0.72 & 0.60 &-0.83 &4.41 \\
    $\pmax$ & 0.68 & 1 & 0.97 & 2.01 & - & - \\
    \hline
    $\pmin$ & 0.90 & 1.64 & 0.88 & 0.68 & 2.03 & -0.76 \\
    $\pmax$ & 0.90 & 1.64 & 0.31 & 2.25 & - & - \\
    \hline
    $\pmin$ & 0.95 & 1.95 &  0.56 & 0.48 & 1.79 & -1.03 \\
    $\pmax$ & 0.95 & 1.95 &  0.22 & 2.54 & - & -\\
    \hline
  \end{tabular}
\caption{\label{tab:clparams} Parameters of the analytic
  approximation to the $\alpha$ level confidence intervals
  \refeq{approxcls} for the MAS normalized estimator.}
\end{table}

\section{The case of a general covariance matrix}
\label{sec:general}

We address now the issue of generalizing the MAS estimator 
to any Stokes parameters covariance matrix. We however consider that the intensity
measurement $I$ is essentially decoupled from $(Q,U)$, as is generally the case in
real-life experiments, and therefore only consider the bi-variate $[q,u]$ covariance matrix.

\subsection{Noise-bias and variance}
\label{sec:nocor}

We ask the following question: for a general $[q,u]$
covariance matrix, what
are the asymptotic equivalents of the noise-bias and variance of the
$p=\sqrt{q^2+u^2}$ distribution?

\begin{figure}
  \centering
  \includegraphics[width=.45\textwidth]{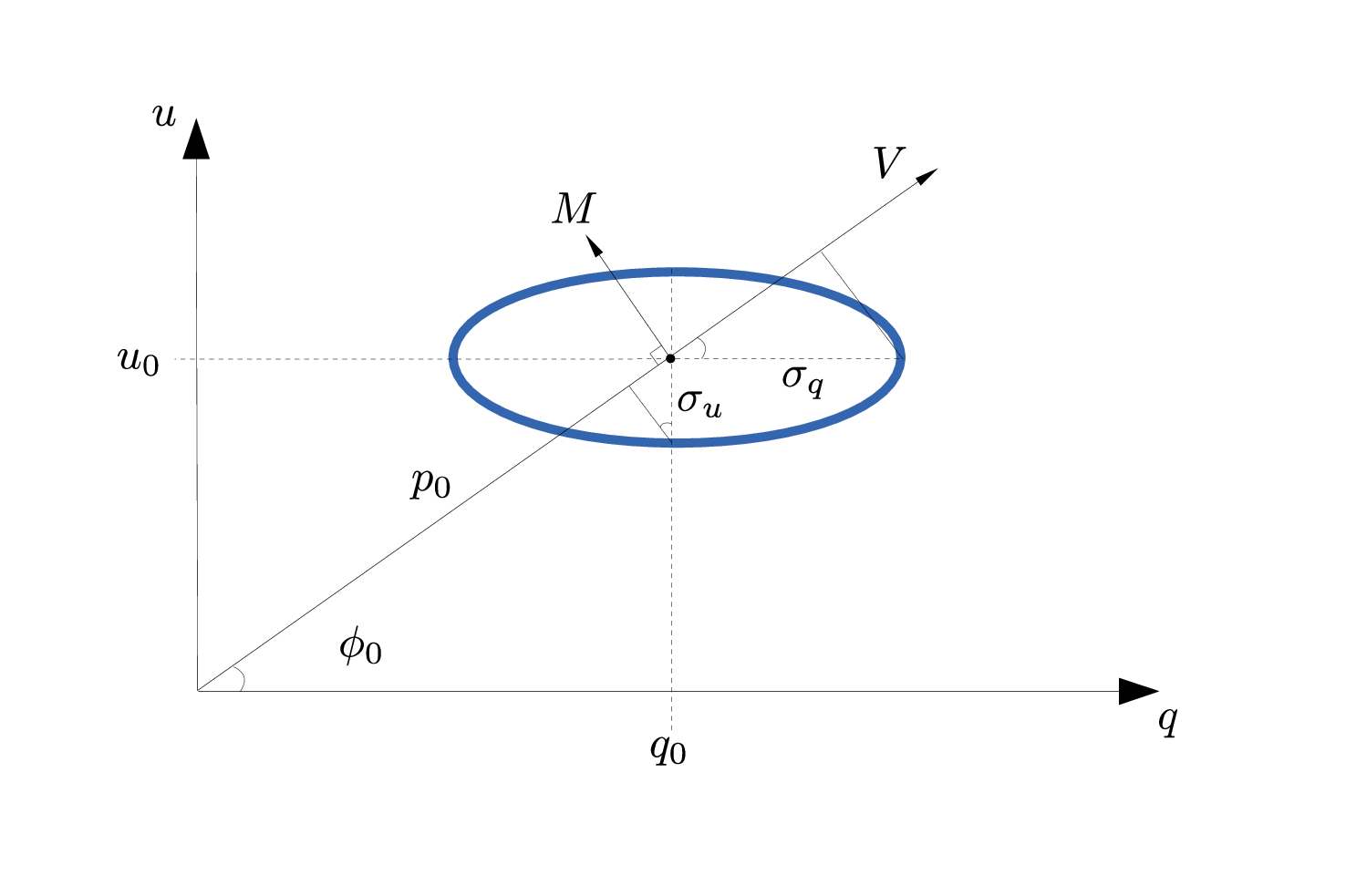}
\caption{\label{fig:geo_ellipse} Same construction as on 
  Fig.~\ref{fig:geo_simple} in the (uncorrelated) elliptical case, $\sigq \ne
\sigu, \rho=0$. The ellipse denotes the 1-$\sigma$ iso-probability $(q,u)$ contour and
one considers the distribution of the distance to the origin of points
located on it. The variance is computed along the centre direction
and gets some contribution from the $\sigma_q\cos\phi_0$ and $\sigma_u\sin\phi_0$  projections,
while the noise-bias has contributions from the orthogonal
combinations $\sigma_q\sin\phi_0$ and $\sigma_u\cos\phi_0$.
In the correlated case, one just needs to rotate the ellipse by
the $\theta$ (\refeq{theta})
angle and re-compute the semi-axis lengths (\refeq{sigrot}).
}
\end{figure}

In the uncorrelated case $(\rho=0$), we formally demonstrate in
Appendix \ref{app:B} that the first two $p$ moments in  the asymptotic regime give:
\begin{align}
\label{eq:b2theo}
\E{p}&=p_0+\dfrac{b^2}{2p_0}; \quad b^2=\sigu^2\cos^2\phi_0+\sigq^2\sin^2\phi_0,\\
\sigma^2_p&=\sigq^2\cos^2\phi_0+\sigu^2\sin^2\phi_0.
\end{align}
Unlike in the canonical case (see \refeq{mean} and \refeq{variance})
the non-linear \bsq{noise-bias} $b^2$ is now different from the variance.
As in the canonical case, these formulas can be understood using the 
simple geometric construction of Fig.~\ref{fig:geo_ellipse}.

We do not know what the true $\phi_0$ angle is. We can either
marginalize over this unknown angle or estimate it for each sample.
If we marginalize over the unknown angle $\phi_0$, we obtain 
the \textit{variance arithmetic mean} for both the noise-bias and the variance:
\begin{equation}
\label{eq:ari}
\sigma_a^2=\half(\sigq^2+\sigu^2).
\end{equation}
In the second approach, we use the fact that
$\phi_i=\atan\tfrac{u_i}{q_i}$ is an asymptotically unbiased estimator
of the angle, even in the elliptical case, and replace the true
angle by it to obtain the \textit{variable bias}:
\begin{align}
  \label{eq:varbias}
  b_i^2& =\sigu^2\cos^2\phi_i+\sigq^2\sin^2\phi_i, \nonumber \\
  &=\dfrac{q_i^2\sigma_u^2+u_i^2\sigma_q^2}{q_i^2+u_i^2},
\end{align}
and similarly the \textit{variable variance}:
\begin{align}
  \label{eq:varvar}
  \sigma_i^2&=\sigq^2\cos^2\phi_i+\sigu^2\sin^2\phi_i \nonumber \\
  &=\dfrac{u_i^2\sigma_u^2+q_i^2\sigma_q^2}{q_i^2+u_i^2}.
\end{align}

In the presence of a non-null correlation coefficient $\rho$ (and
$\sigma_q\ne\sigma_u$), the principal axes of the iso-probability
ellipse are rotated by the angle (\eg \cite*{Aalo2007}):
\begin{equation}
  \label{eq:theta}
  \theta=\half\atan\dfrac{2\rho\sigq\sigu}{\sigq^2-\sigu^2},
\end{equation}
and the semi-diameters along the principal axes are the eigenvalues of
the covariance matrix:
\begin{equation}
  \label{eq:sigrot}
  \begin{split}
  \sigq^{\prime2}&=\sigq^2\cos^2\theta+\sigu^2\sin^2\theta+\rho\sigq\sigu\sin2\theta, \\
  \sigu^{\prime2}&=\sigq^2\sin^2\theta+\sigu^2\cos^2\theta-\rho\sigq\sigu\sin2\theta.
  \end{split}
\end{equation}

Relying on Fig.~\ref{fig:geo_ellipse}, the variance is computed along
the $\phi_0$ direction and the bias along the orthogonal one, in
that case after a rotation of the principal axes by $\theta$.
This result can also be established more formally using computations
along the lines of Appendix \ref{app:B}, by diagonalizing the covariance matrix in the
exponential argument of the Gaussian.
The results depend however very loosely on the correlation
value, since $\sigq^{\prime}~(\sigu^{\prime})$ represents also essentially
a rotation of $\sigq~(\sigu^{\prime})$. For values of $\rho
\lesssim 0.5$ one can safely neglect it and use the previous results.

The marginalized result with a correlation gives back the variance arithmetic
mean since:
\begin{equation}
  \label{eq:aricor}
  \half (\sigq^{\prime2}+\sigu^{\prime2})=\half(\sigq^2+\sigu^2)=\sigma_a^2,
\end{equation}
and the variable estimates from $\phi_i=\atan\dfrac{u_i}{q_i}$ is:
\begin{align}
  \label{eq:varbiasrho}
  b_i^2&=\sigu^{\prime2}\cos^2(\phi_i-\theta)+ \sigq^{\prime2}\sin^2(\phi_i-\theta),\\
  \label{eq:varvarrho}
  \sigma_i^2&=\sigq^{\prime2}\cos^2(\phi_i-\theta)+ \sigu^{\prime2}\sin^2(\phi_i-\theta).
\end{align}

\begin{figure}
  \centering
  \includegraphics[width=.5\textwidth]{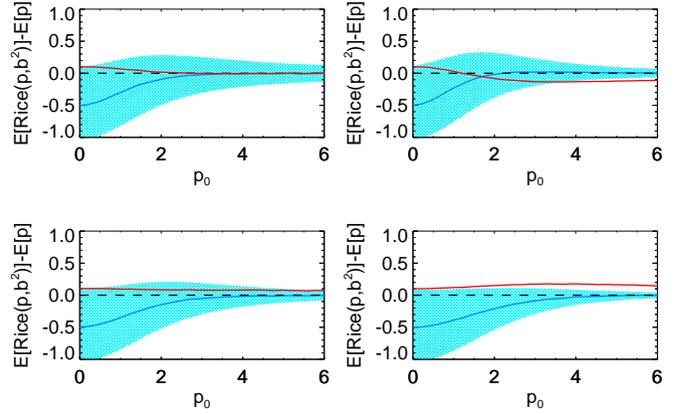}
\caption{\label{fig:mom1_cor} Validation of the Rice equivalent
  noise-bias in the 
  elliptic case $\sigma_q=1,\sigma_u=2, \rho=0.7$, for
  several polarization angles: upper
  left for a uniform angle distribution, upper right for $\phi_0=0\deg$,
  lower left $\phi_0=40\deg$, lower right $\phi_0=80\deg$. 
  In each case, the expectation value of the complete distribution
  $E[p]$ is obtained from Monte-Carlo simulation and is compared to
  the Rice expectation value (\refeq{mean}) using for the $\sigma$
  term, 
  either the variance arithmetic mean (red line, \refeq{ari})
  or the mean of the variable noise estimate (blue line, \refeq{varbiasrho}). 
  In this latter case the shaded blue region shows the $1\sigma$
  variation of the estimates.
} 
\end{figure}

\begin{figure}
  \centering
  \includegraphics[width=.5\textwidth]{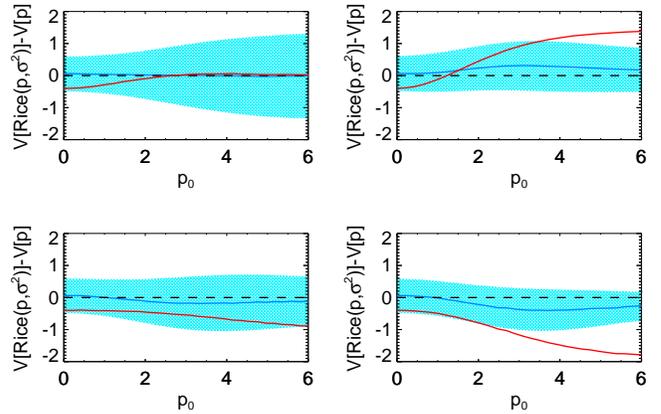}
\caption{ \label{fig:mom2_cor} Same as Fig.~\ref{fig:mom1_cor} but for
  the Rice equivalent variance in the same elliptic case $\sigma_q=1,\sigma_u=2, \rho=0.7$.
  Upper left is for a uniform angle distribution, upper right for $\phi_0=0\deg$,
  lower left $\phi_0=40\deg$, lower right $\phi_0=80\deg$.
  We compare the empirical variance $V[p]$, obtained from Monte-Carlo
  simulations, 
  to the variance of the Rice distribution (\refeq{variance}) using for  
  the $\sigma$ term, either the variance arithmetic mean (red line, \refeq{ari}),
  or the mean of the variable noise estimate (blue line, \refeq{varvarrho}).
  In this latter case the shaded blue region shows the $1\sigma$
  variation of the estimates.}
\end{figure}

We test the validity of these estimates in a highly elliptic and
correlated case $\sigq=1,\sigu=2,\rho=0.7$. Results are presented on Fig.~\ref{fig:mom1_cor}
for the bias and  Fig.~\ref{fig:mom2_cor} for the variance, for
several true polarization angles.

The variable noise-bias is found to match very precisely and rapidly the
empirical expectation value, while the variance arithmetic mean may
slightly over- or under-estimate the asymptotic values, depending on the underlying true
angle. For the Rice-equivalent variance,
the variable variance is reasonable in the whole $p_0$ range, while
the arithmetic mean may lead to a severe asymptotic discrepancy for 
some angles.

\subsection{Generalized MAS estimator}

Since our aim is to build an estimator which is unbiased as fast as
possible with the SNR, we generalize the MAS estimator to: 
\begin{equation}
  \label{eq:pmas2}
    \pmas=p_i- b_i^2\frac{1-e^{-p_i^2 /b_i^2}}{2p_i},
\end{equation}
where the noise-bias $b_i$ is computed on a sample by sample basis, either from
\refeq{varbias} for the uncorrelated case, or \refeq{varbiasrho} for the
(strongly) correlated one.

\begin{figure}
  \centering
  \includegraphics[width=.5\textwidth]{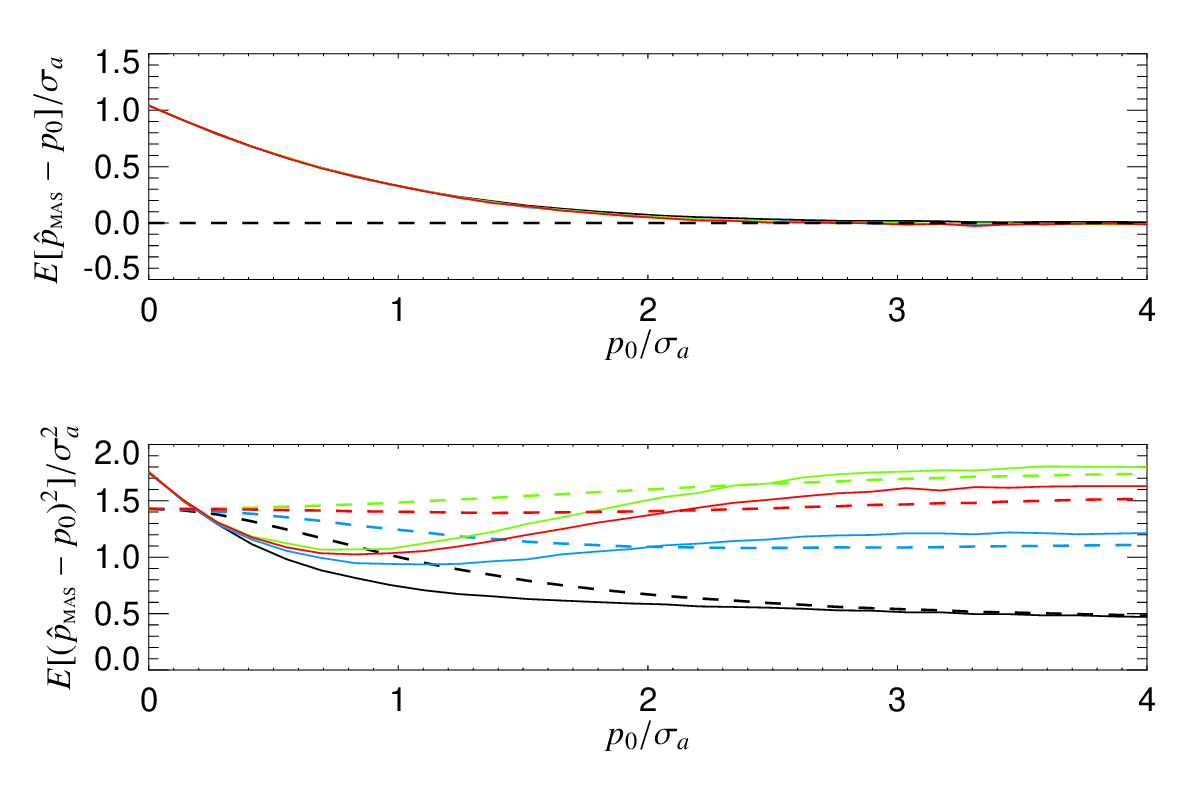}
  \caption{\label{fig:bias_risk_scan} Bias and risk, normalized by the
    variance arithmetic mean $\sigma_a$, of the generalized
    MAS estimator in the $\sigq=1,\sigu=2,\rho=0.7$ case, as obtained
    from Monte-Carlo simulations. The true $\phi_0$ angle is varied
    according to the following color code: 0\deg(black), 30\deg(blue),
  60\deg(red), 90\deg(green). The dashed lines show the variance
  estimates $\sigma_i^2$ from \refeq{varvarrho}.}
\end{figure}

We re-consider its bias and risk on Fig.~\ref{fig:bias_risk_scan} in
the  highly elliptic regime, for several $\phi_0$ angles.
As may have been anticipated from the previous section results, even
in this rather extreme case, the bias is insensitive to the true angle and
is very similar to the canonical case, \ie essentially unbiased
above 2.
The risk depends now on the true angles, but since the estimator has no
bias in this region, its risk is equivalent to its variance and
our \refeq{varvarrho} estimate provides a reasonable asymptotic description.

%%%%%%%%%%%%%%%%%%%%%%%%%%%%%%%%%%%%%%%%%%%%%%%%%%%%%%%%%%%%%%%%%
\section{Conclusion}
\label{sec:conclusion}

We have developed and characterized an estimator of the polarization amplitude that enjoys
several desirable properties. 
Its distribution lies in the positive region, is continuous, and
transforms smoothly with the SNR from a Rayleigh-like to a Gaussian one,
 the latter being essentially reached above 2.  

We revisited the construction of confidence intervals and solved
efficiently the empty-set (or unphysical) region problem encountered
at low SNR using the Feldman-Cousins prescription. We provided
analytic approximations to the 0.68, 0.90 and 0.95 confidence level
regions. 

We have generalized the estimator to the case of a global covariance
matrix, and shown that its bias is universal, \ie independent of the
true $\phi_0$ angle. We provided an analytic estimate of the
variance of the estimator that can be used to assess the risk in the
large SNR region.

Given its very simple analytic form, the estimator can be applied
efficiently on large data-sets, in particular for providing
Gaussian-like point-estimates in regions of reasonably large SNR values, and
conversely build masks to identify regions not bearing enough statistical
significance.
This can be performed using the following procedure:
\begin{enumerate}
\item compute \pmas from \refeq{pmas2} and the variance arithmetic mean
  $\sigma_a$ from \refeq{ari} from all data pixels.
\item according to Sect. \ref{sec:cl} results, a SNR above 2 at the 90\% CL
  is obtained by keeping samples satisfying $\tfrac{\pmas}{\sigma_a} >
  3.8$. 
 This is used to build a mask, that can possibly be spatially smoothed.
\item in the rest of the data, point-estimates can be given safely
  since we have shown that in this regime the estimator is unbiased
  and essentially Gaussian. One can compute the estimator variance
  using \refeq{varvarrho} and consider it as its associated \bsq{error}.
\end{enumerate}
For values within the mask, reporting a point-estimate is unsafe and one
should instead report a confidence interval, as 
the ones given in Sect. \ref{sec:cl}, or a full likelihood function.

This work was oriented towards estimating the polarization
amplitude but is obviously much more general. It is perhaps surprising
that such a fundamental question as characterizing the amplitude of a
vector or the modulus of a complex number from its normally distributed Cartesian components did not receive more
attention.
A part of the reason is maybe related to defining precisely the
question: what is a \bsq{good} estimator?  We tried to answer it in a
user-oriented way.

\section*{Acknowledgments}
We thank Jason L. Quinn for an efficient and in-depth refereeing of the manuscript.
%%%%%%%%%%%%%%%%%%

% BIBTEX
\bibliographystyle{mn2e}
\bibliography{refs_aps}

\begin{thebibliography}{}

\bibitem[\protect\citeauthoryear{{Aalo}, {Efthymoglou} \& {Chayawan}}{{Aalo}
  et~al.}{2007}]{Aalo2007}
{Aalo} V.~A.,  {Efthymoglou} G.~P.,    {Chayawan} C.,  2007, IEEE
  Communications Letters, 11, 985

\bibitem[\protect\citeauthoryear{{Beringer~\etal}}{{Beringer~\etal}}{2012}]{PDG2012}
{Beringer~\etal} J.,  2012, Phys. Rev. D, 86, 010001

\bibitem[\protect\citeauthoryear{{C\'ardenas-Blanco} A.~{Tejos} \&
  {Cameron}}{{C\'ardenas-Blanco} \& {Cameron}}{2008}]{Cardenas2008}
{C\'ardenas-Blanco} A.~{Tejos} C. I.~P.,  {Cameron} I.,  2008, Concepts Magn.
  Reson., pp 409--416

\bibitem[\protect\citeauthoryear{Chandrasekar}{Chandrasekar}{1950}]{Chandrasekar1950}
Chandrasekar S.,  1950, Radiative Transfer.
Oxford Univ. Press

\bibitem[\protect\citeauthoryear{{Feldman} \& {Cousins}}{{Feldman} \&
  {Cousins}}{1998}]{FC}
{Feldman} G.~J.,  {Cousins} R.~D.,  1998, \prd, 57, 3873

\bibitem[\protect\citeauthoryear{{Gradshteyn} \& {Ryzhik}}{{Gradshteyn} \&
  {Ryzhik}}{2007}]{GradshteynRyzhik2007}
{Gradshteyn} I.~S.,  {Ryzhik} I.~M.,  2007, {Table of Integrals, Series, and
  Products}.
Academic Press

\bibitem[\protect\citeauthoryear{{Gudbjartsson} \& {Patz}}{{Gudbjartsson} \&
  {Patz}}{1995}]{Gudbjartsson1995}
{Gudbjartsson} H.,  {Patz} S.,  1995, Magn Reson Med, 34, 910

\bibitem[\protect\citeauthoryear{{James}}{{James}}{2007}]{James2007}
{James} F.,  2007, Statistical Methods in Experimental Physics.
World Scientific

\bibitem[\protect\citeauthoryear{{Neyman}}{{Neyman}}{1937}]{Neyman1937}
{Neyman} J.,  1937, Phil. Trans. Royal Soc. London Ser., A, 333

\bibitem[\protect\citeauthoryear{Olver, Lozier, Boisvert \& Clark}{Olver
  et~al.}{2010}]{NIST2010}
Olver F.~W.~J.,  Lozier D.~W.,  Boisvert R.~F.,    Clark C.~W.,  eds, 2010,
  {NIST Handbook of Mathematical Functions}.
Cambridge University Press, New York, NY

\bibitem[\protect\citeauthoryear{{Quinn}}{{Quinn}}{2012}]{Quinn2012}
{Quinn} J.~L.,  2012, \aap, 538, A65

\bibitem[\protect\citeauthoryear{{Rice}}{{Rice}}{1945}]{Rice1945}
{Rice} S.~O.,  1945, Bell Systems Tech.~J., 24, 46

\bibitem[\protect\citeauthoryear{Sijbers}{Sijbers}{1998}]{SijbersThesis1998}
Sijbers J.,  1998, PhD thesis, Universiteit Antwerpen

\bibitem[\protect\citeauthoryear{Sijbers, den Dekker, Scheunders \&
  Dyck}{Sijbers et~al.}{1998}]{Sijbers1998}
Sijbers J.,  den Dekker A.,  Scheunders P.,    Dyck D.~V.,  1998, IEEE
  Transactions on Medical Imaging, 17, 357

\bibitem[\protect\citeauthoryear{{Simmons} \& {Stewart}}{{Simmons} \&
  {Stewart}}{1985}]{Simmons1985}
{Simmons} J.~F.~L.,  {Stewart} B.~G.,  1985, \aap, 142, 100

\bibitem[\protect\citeauthoryear{{Talukdar} \& {Lawing}}{{Talukdar} \&
  {Lawing}}{1991}]{Taludkar1991}
{Talukdar} K.~K.,  {Lawing} W.~D.,  1991, Journal of the Acoustical Society of
  America, 89, 1193

\bibitem[\protect\citeauthoryear{{Vinokur}}{{Vinokur}}{1965}]{Vinokur1965}
{Vinokur} M.,  1965, Annales d'Astrophysique, 28, 412

\bibitem[\protect\citeauthoryear{{Wardle} \& {Kronberg}}{{Wardle} \&
  {Kronberg}}{1974}]{Wardle1974}
{Wardle} J.~F.~C.,  {Kronberg} P.~P.,  1974, \apj, 194, 249

\end{thebibliography}

%%%%%%%%%%%%%%%%%
\appendix

%%%%%%%%%%%%%%%%%%%%%%
\section{Bias of any transform of the Rayleigh distribution}
\label{app:A}

We demonstrate in this appendix that one cannot build an estimator
that completely removes the bias on the $p=\sqrt{q^2+u^2}$ variable.
For the sake of simplicity, we work in the canonical frame ($\sigq=\sigu=1,
\rho=0$) with a true amplitude value of $p_0=0$. The random variable
then follows the Rayleigh distribution:
\begin{equation}
  \label{eq:rayleigh}
  f_p(p)=p e^{-p^2/2}.
\end{equation}
We then ask the following question: can we find a change of variable
for which the resultant distribution would be completely unbiased,
\ie have a mean of 0?

Let us consider \textit{any} (bijective) transformation $f$: 
\begin{align}
    \hat p &= f(p), \\
    p &= f^{-1}(\hat p)\equiv g(\hat p).
\end{align}
The transformed probability density is:
\begin{align}
    f_{\hat p}(\hat p) &=g(\hat p)g^\prime(\hat p) e ^{-g(\hat p)^2/2} \\
    &=-\dfrac{d}{d \hat p}(e^{-g(\hat p)^2/2}).
\end{align}
The characteristic function of $H(\hat p)\equiv e^{-g(\hat p)^2/2}$
being 
\begin{equation}
\phi(k)=\E{e^{ik\hat p}}=\int_0^\infty d \hat p H(\hat p) e^{ik\hat p},
\end{equation}
the characteristic function of ${\hat p}$ is classically:
\begin{equation}
  \Phi(k)=-(-i k \phi(k))=ik\phi(k).
\end{equation}
Taking its first order derivative:
\begin{equation}
  \Phi^\prime(k)=i \phi(k) +i k \phi^\prime(k).
\end{equation}
The mean of ${\hat p}$ is then given by 
\begin{equation}
  \E{\hat p}= \dfrac{1}{i}\Phi^\prime(0)=\phi(0)=\int d\hat p H(\hat p) =\int d\hat p
  e^{-g^2(\hat p)/2}.
\end{equation}

Whatever the initial transform is, the integrand is always positive and
the estimator is\textit{ always positively biased}.  This can be
decreased by choosing a rapidly decaying function  (as $\hat p=\log
p$) but at the price of introducing some negative values.

\section{Computation of the noise-bias and variance in the elliptical case}
\label{app:B}

The normal probability density of the uncorrelated $(q,u)$ bi-variate
variable is \footnote{We indicate in Sect.~\ref{sec:nocor} how to
  perform the computation if the correlation coefficient $\rho$ does
  not equal zero.}:
\begin{equation}
  f_{q,u}(q,u)=\dfrac{1}{2\pi\sigma_q\sigma_u}e^{-\left(\dfrac{(q-q_0)^2}{2\sigq^2}+\dfrac{(u-u_0)^2}{2\sigu^2}\right)},
\end{equation}
and its first-order moment is computed from:
\begin{equation}
  \E{p}=\intinf \intinf dq du \sqrt{q^2+u^2} f_{q,u}(q,u).
\end{equation}
We apply the change of variables $\bar q=\tfrac{q-q_0}{\sigq}, \bar u=\tfrac{u-u_0}{\sigu}$ to obtain: 
\begin{equation}
  \label{eq:ep}
  \begin{split}
  \E{p}&=\iint d\bar q d\bar u \dfrac{1}{2\pi} e^{-\dfrac{\bar q^2+\bar u^2}{2}} \\
  &\qquad \times \sqrt{(\sigq \bar q+q_0)^2+(\sigu \bar u+u_0)^2}
  \end{split}
 \end{equation}
The square-root term can be expressed as: 
\begin{align}
  &p_0\left[ 1+ 2 c_0 \dfrac{\sigq}{p_0} \bar q +2
    s_0\dfrac{\sigu}{p_0}  \bar u + \left(\dfrac{\sigq}{p_0}\right)^2 \bar q^2
      + \left(\dfrac{\sigu}{p_0}\right)^2 \bar u^2 \right]^{1/2} \\
    &\equiv p_0\sqrt{1+x} \nonumber
\end{align}
where we introduced the shorthand notation $c_0\equiv\cos\phi_0$ and
$s_0\equiv\sin\phi_0$, with $\phi_0$ as the true polar angle.

The product with the Gaussian function in \refeq{ep} restricts the sizable range in 
the integral to about $|\bar q| \lesssim2$ and $|\bar u|\lesssim2$.
For a high SNR $\epsilon_q\equiv\dfrac{\sigq}{p_0}\ll\dfrac{1}{2}$ and
$\epsilon_u\equiv\dfrac{\sigu}{p_0}\ll\dfrac{1}{2}$, so that finally $|x|<1$. 
We then perform the series expansion of $\sqrt{1+x}$ in $x$ at $x=0$
omitting odd powers of $\bar q$ and $\bar u$, since
their further product with the Gaussian cancels in the integral
according to the parity relation:
\begin{equation}
  \intinf\intinf d\bar q du~ \bar q^{2k+1} e^{-\dfrac{\bar q^2+\bar u^2}{2}}=0,
\end{equation}
for any $k$ integer, and similarly for the $\bar u^{2k+1}$ terms.
The remaining leading terms are: 
\begin{align}
  &1 +\dfrac{1}{2}(\epsilon_q^2 \bar q^2+\epsilon_u^2 \bar u^2)
  -\dfrac{1}{8}(4\epsilon_q^2 c_0^2 \bar q^2 + 4\epsilon_u^2 s_0^2
  \bar u^2 ) +\bigO{\epsilon^4} \nonumber \\
&=1 + \dfrac{1}{2}(\epsilon_q^2 s_0^2 \bar q^2+ \epsilon_u^2 c_0^2 \bar
u^2) +\bigO{\epsilon^4}.
\end{align}
Making use of 
\begin{equation}
  \dfrac{1}{2\pi}\intinf\intinf d\bar q d\bar u ~\bar q^2 e^{-\dfrac{\bar
      q^2+\bar u^2}{2}} =1,
\end{equation}
and similarly for $\bar u^2$, we finally obtain: 
\begin{align}
  \label{eq:meanEP}
  \E{p}&=p_0\left[ 1+ \dfrac{1}{2}(\epsilon_q^2 s_0^2 + \epsilon_u^2
    c_0^2) \right]  \nonumber \\
  &=p_0+\dfrac{\sigu^2\cos^2\phi_0+\sigq^2\sin^2\phi_0}{2p_0}.
\end{align}

By comparing this expression to \refeq{mean} we see that the
equivalent noise bias that takes into account ellipticity is given by:
\begin{equation}
  b^2=\sigu^2\cos^2\phi_0+\sigq^2\sin^2\phi_0,
\end{equation}
which gives indeed back $\sigma^2$ in the canonical case.

For the variance, a similar computation leads to:
\begin{align}
  \E{p^2}&=\intinf \intinf dq du (q^2+u^2) f_{q,u}(q,u) \nonumber\\
  &=p_0^2+\sigq^2+\sigu^2.
\end{align}
Using \refeq{meanEP} and expanding $\E{p}^2$ keeping first order terms:
\begin{align}
    V&=\E{p^2}-\E{p}^2 \nonumber \\
    &=(p_0^2+\sigq^2+\sigu^2)-(p_0^2+\sigu^2\cos^2\phi_0+\sigq^2\sin^2\phi_0) \nonumber \\
    &=\sigq^2\cos^2\phi_0+\sigu^2\sin^2\phi_0.
\end{align}

\end{document}